\documentclass[preprint,showpacs,preprintnumbers,showkeys,amsmath,amssymb]{revtex4}

\usepackage{graphicx}
\usepackage{dcolumn}
\usepackage{bm}

\begin{document}


\title{Three-dimensional Negative-Refractive-Index Metamaterials Based on All-Dielectric Coated Spheres}

\author{Na Fu}
\author{Weixing Shu}\thanks{Corresponding author. $E$-$mail$ $address$: wxshuz@gmail.com.}
\author{Hailu Luo}
\author{Zhixiang Tang}
\author{Yanhong Zou}
\author{Shuangchun Wen}
\author{Dianyuan Fan}
\affiliation{ Key Laboratory for Micro/Nano Optoelectronic Devices
of Ministry of Education, School of Computer and Communications,
Hunan University, Changsha 410082, China}

\begin{abstract}
A type of 3-dimensional  optical negative-refractive-index
metamaterials composed of all dielectric nanospheres is proposed and
demonstrated theoretically. The metamaterials are constructeded by
pairing together two kinds of dielectric nanospheres as concentric
shells embedded in a host medium. Mie-based extended effective
theory shows that the dielectric core and the dielectric shell
provide the negative permeability and the negative permittivity,
respectively, both due to the strong Mie resonances. Within the
coupled resonant frequency region, the negative index of refraction
can be achieved.
\end{abstract}


\maketitle

\section{Introduction}\label{Introduction}

Negative-index metamaterials(NIMs) \cite{Veselago1968} are
artificial structures that exhibit negative index of refraction.
Since these metamaterials enable a variety of novel applications
such as superlens \cite{Tsakmakidis2007}, optical nanocircuits
\cite{Engheta2007}, cloaking \cite{Pendry2006,Alu2007}, marvellous
progresses on the NIMs have been made
\cite{Shelby2001,Pendry2000,Smith2004} in recent years.

Along with the rapid development of NIMs, it has been challenging to
design 3D isotropic NIMs at optical frequencies. So far, efforts
devoted to designing NIMs at optical frequencies mainly consist of
two basic ideas, namely those based on L-C resonant models and the
Mie scattering models. The most prominent structures based on L-C
models are metal-dielectric fishnet structures \cite{Valentine2008}.
However, due to the anisotropy of these metallic structures, the
negative refraction can only be performed in one certain direction.
Another drawback of the metal based structures is the difficulty in
enhancing resonant frequencies and fabrication. The Mie scattering
model based structures are those  multilayered metal and dielectric
microspheres\cite{Yannopapas2007}, metal coated dielectric spheres
\cite{Wheeler2006}, et al. Also, there are some problems, such as
large losses and saturation effect, inherent in metals at optical
frequencies associated with these metallic metamaterials. As an
alternative, the Mie resonance of highly polaritonic dielectric
materials, such as rod \cite{Peng2007,Schuller2007,Vynck2009} or
cube type \cite{Zhao2008} metamaterials, provides a more promising
way to design low-loss, much simpler 3D metamaterials with higher
frequencies. However, the ultimate goal of realizing 3D
metamaterials at optical frequencies has not been fulfilled up to
now.

  In this paper, we theoretically propose a type of optical negative refractive index
metamaterial that composed of all dielectric coated nanospheres. Our
objective is to use low-permittivity dielectric materials to
increase the electromagnetic resonant frequencies to optical domain,
and to utilize highly symmetrical sphere-type structure to realize
3D optical NIMs. For dielectric nanospheres, a strong magnetic
dipole resonance results in the negative effective permeability. And
an electric dipole resonance leads to the negative effective
permittivity when the dielectric constant and radius of the spheres
are increased. We proceed as follows. Firstly, we use Mie theory to
describe the effective magnetic and electric resonances, and then
derive the relationship between the effective permeability and
permittivity and the material parameters by homogenizing the
spherical scatters. Secondly, we show that the negative magnetic and
electric response can be produced at optical frequencies by tuning
the sphere parameters. Lastly, we tune these resonances together
using coated dielectric spheres and the negative refractive index
can be obtained. Theoretically, a negative index of refraction can
be obtained at any optical frequency.

\section{Negative magnetic and electric response by Mie-based Maxwell-Garnett
theory}\label{II}

The theoretical basis of our study is the Mie-based
Maxwell-Garnett(MMG) theory \cite{Ruppin2000}. We consider a
composite of small dielectric spheres of radius $a$ and dielectric
constant $\epsilon_r$ embedded in a host medium with dielectric
constant $\epsilon_h$ and incident wavelength $\lambda_h$. Generally
Mie scattering can be used when
$(0.05\sim0.1)\lambda_h<2a<(3\sim6)\lambda_h$ \cite{van1964}. Let
$x$ be the size parameter which can be defined as $x=2\pi a
/\lambda_h $. If $x\ll1$, for a periodic distribution, the effective
permeability ${\bf \mu}_{eff}$ is given by Clausius-Mossotti
equation,
\begin{eqnarray}\label{a}
\frac{{\bf \mu}_r^{eff}-1}{{\bf \mu}_r^{eff}+2}=\frac{f}{a^3}\alpha,
\label{a}
\end{eqnarray}
where $\alpha$ is the particle dipole polarizability, $f=4\pi N
a^3/3$ is the filling fraction of the composite, and $N$ is the
number density of the spheres.

  As long as $x$ is small enough so that the Mie coefficients $a_m$ and
$b_m$ with $m>1$ can be neglected, a size dependent extension of the
MG formula \cite{Doyle1989} suggests that in terms of the Mie
coefficient the polarizability is
\begin{eqnarray}\label{h_+/e_+}
{\bf \alpha}=i\frac{3a^3}{2x^3}b_1, \label{b}
\end{eqnarray}
and the Mie scattering coefficients, $a_m$ and $b_m$ are
\begin{eqnarray}\label{c}
a_m=\frac{n{\bf \psi}_m(nx){\bf \psi}'_m(x)-{\bf \psi}'_m(nx){\bf
\psi}_m(x)}{n{\bf \psi}_m(nx){\bf \xi}'_m(x)- {\bf \psi}'_m(nx){\bf
\xi}_m(x)}, \label{c}
\end{eqnarray}
\begin{eqnarray}\label{d}
b_m=\frac{{\bf \psi}_m(nx){\bf \psi}'_m(x)-n{\bf \psi}'_m(nx){\bf
\psi}_m(x)}{{\bf \psi}_m(nx)'{\bf \xi}_m(x)-n{\bf \psi}'_m(nx){\bf
\xi}_m(x)}, \label{d}
\end{eqnarray}
respectively, where $\psi_m(x) $ and $\xi_m(x) $ are related to the
Riccati-Bessel functions.

   Combined with Eqs.~(\ref{a}) and (\ref{b}), the effective permeability for
a homogenizing distribution of inclusions can be obtained as
\begin{eqnarray}\label{e}
{\bf \mu}_r^{eff}={\bf \epsilon}_h(1+\frac{6\pi iN_\mu
b_1}{\epsilon_h^3k^3-2\pi iN_\mu b_1}), \label{e}
\end{eqnarray}
where $k=\omega_\mu^{inc}/c$, and $\omega_\mu^{inc}$ is the
frequency of the incident wave. Analogously, the effective
permittivity ${\bf \epsilon}_r^{eff}$ can be expressed as
\begin{eqnarray}\label{f}
{\bf \epsilon}_r^{eff}={\bf \epsilon}_h(1+\frac{6\pi iN_\epsilon
a_1}{\epsilon_h^3k^3-2\pi iN_\epsilon a_1}). \label{f}
\end{eqnarray}
Eqs.~(\ref{e}) and (\ref{f}) imply that the effective permeability
and permittivity depend on the frequency of the incident wave, the
host medium, Mie coefficient as well as the number density of the
spheres.

\subsection{Magnetic response}
  Eq.~(\ref{e}) reveals the magnetic
resonant frequency $\omega_\mu^{res}$ occurs if the frequency of the
incident wave satisfy
\begin{eqnarray}\label{g}
\omega_\mu^{inc}=\frac{c}{\epsilon_h}(2 \pi i N_\mu b_1)^{1/3}.
\label{g}
\end{eqnarray}
Eq.~(\ref{g}) implies the magnetic response requires appreciable
values of the $N_\mu b_1$. For a moderate filling fraction, the
magnetic resonance mainly results from the fundamental Mie resonance
\cite{Vynck2009}. The scattering properties of the sphere-type
structures may therefore be understood by studying the resonant
behavior of $b_1$. Using the half-integer Bessel function, ${\bf
\psi}_m(x)=\sqrt{\pi x/2}J_{m+1/2}(x) $ and ${\bf
\xi}_m(x)=\sqrt{\pi x/2} [J_{m+1/2}(x)+(-1)^m J_{-m-1/2}(x)]$, for
$m=1$, the simplified resonance occurs when
$J_0(nka)=sin(nka)/(nka)=0$ \cite{Wheeler2006}. The fundamental Mie
resonance frequency is $\omega_\mu^{res}=\pi c/(a
\sqrt{\epsilon_r\epsilon_h})$. We assume, as is typical for
dielectric resonators, that the resonance frequencies are determined
by the real part of the refractive index of the sphere
$n=\sqrt\epsilon_r$.
   The relative permittivity of polaritonic dielectrics follows
$\epsilon_r=\epsilon_\infty[1+(\omega_L^{2}-\omega_T^{2})/(\omega_T^2-\omega^2-i\omega\gamma)]$
where $ \epsilon_\infty$ is the high frequency limit of the
dielectric permittivity, $\gamma$ is the loss factor, and $\omega_T$
and $\omega_L$ are the transverse and longitudinal optical phonon
frequencies. Since the electromagnetic resonant frequency region we
considered is much larger than $\omega_T$, so in this paper we take
the approximation of $ \epsilon_r= \epsilon_\infty$.
\begin{figure}
\includegraphics[width=8cm]{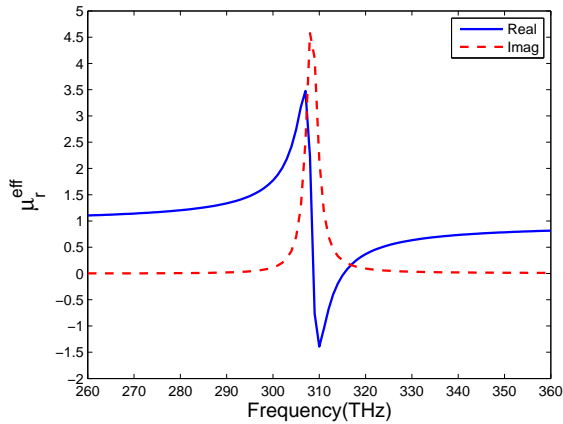}
\caption{\label{FuFig1}Calculated effective permeability $ {\bf
\mu}_{eff}$ for a periodic distribution of dielectric spherical
particles with radius $a=120nm $ and the dielectric constant ${\bf
\epsilon}_r=2.2$. The dielectric constant of the host medium is
${\bf \epsilon}_h=1.0 $ and the number density of the composite is
$N_\mu=(242.6nm)^{-3}$.}
\end{figure}

   Theoretically, almost arbitrary values of $ {\bf \mu}_{eff}$
can be obtained in a collection of appropriate size of non-magnetic
spheres. As an example, consider a collection of dielectric spheres
using the parameters: ${\bf\epsilon}_r=n^2=2.2$ with the number
density $N_\mu=(242.6nm)^{-3}$ and radius $a=120nm $. For
simplicity, we choose host medium as ${\bf\epsilon}_h=1.0$.
According to Eq.~(\ref{g}), the magnetic resonance is predicted at
310THz. A full calculation of Eq.~(\ref{e})(shown in
Fig.~\ref{FuFig1}) reveals a resonance in $\mu_{eff} $ centered at
310THz with a real value of -1.5, which agrees well with our
prediction. Additionally, from Eq.~(\ref{g}) we also find the number
density of spheres $N_\mu$ and the dielectric constant of the host
medium $\epsilon_h $ will affect the magnetic resonant frequency.
This is vigorous to assist the design of a negative refractive index
in Sec. \ref{III}.

\subsection{Electric response}
The method described in deciding the magnetic response can also be
used to calculate $a_1$. From Eq.~(\ref{f}), the dielectric
resonance $\omega_\epsilon^{res}$ is induced when
\begin{eqnarray}\label{h}
\omega_\epsilon^{inc}=\frac{c}{\epsilon_h}(2 \pi i N_\epsilon
a_1)^{1/3}. \label{h}
\end{eqnarray}
The resonant frequencies of $a_1$ in the long wavelength limit can
be estimated by
\begin{eqnarray}\label{i}
\frac{J_0(nka)+J_2(nka)}{J_0(nka)-J_2(nka)}+\frac{1}{\epsilon_r}=0.
\label{i}
\end{eqnarray}
Eq.~(\ref{i}) is only suitable for numerical calculation and the
resonant frequency is related to the dielectric constant of the
nanospheres $\epsilon_r$. Since the first Mie resonance $a_1$ occurs
in lower frequencies than the fundamental Mie resonance $b_1$ with
the same parameters, it requires a dielectric constant larger than
that in Sec. \ref{II} A to drive an equivalent resonance.
\begin{figure}
\includegraphics[width=8cm]{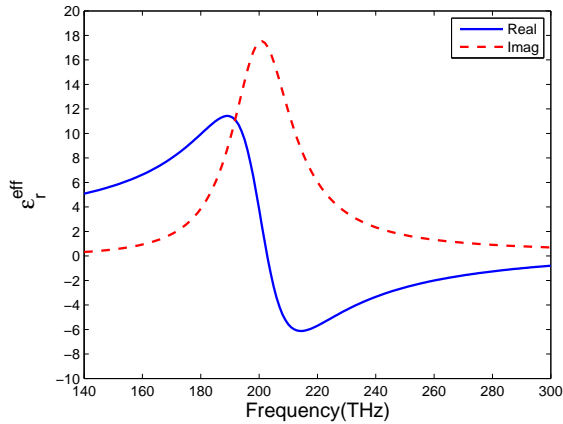}
\caption{\label{FuFig2}Calculated effective permeability $ {\bf
\epsilon}_{eff}$ for a periodic distribution of identical dielectric
spherical particles with radius $a=120nm $ and the dielectric
constant ${\bf \epsilon}_r=13.4$. The dielectric constant of the
host medium is ${\bf \epsilon}_h=1.0 $ and the number density is
$N_\mu=(242.6nm)^{-3}$.}
\end{figure}

   Take the parameter $\epsilon_r=13.4$ of $\text{LiTO}_3$ \cite{Huang2004},
and the same number density as in the above section. The lattices of
$\text{LiTO}_3$ spheres are predicted to possess the fundamental
dielectric response at 212THz. Calculation of Eq.~(\ref{f}) (shown
in Fig.~\ref{FuFig2}) indicates that the first electric resonance is
induced at approximately 210THz with a value of -7. This is
particularly helpful to provide a desired electric response in Sec.
\ref{III}.

\section{Negative index of Refraction}\label{III}
   In the above two section, we have realized
the negative permeability and the negative permittivity separately.
However, a single type of dielectric spheres collection cannot serve
as a negative refractive index metamaterial on its own. In order to
pair together the magnetic resonant frequency and the electric
resonant frequency in the same frequency region, one can require
$N_\mu b_1=N_\epsilon a_1$. In particular, we find that utilizing
the relationship between size parameter $x$ and the Mie resonant
frequency can narrow the gap between $a_1$ and $b_1$. Concentric
dielectric nanospheres enable the problem to be solved and provide a
novel mechanism for the creation of 3D negative index metamaterial.
We can choose dielectric sphere as the core to provide a desired
magnetic resonance, and larger radius shell with higher dielectric
constant material cover on the core to increase the electric
resonant frequencies to corresponding region. When these two
resonances simultaneously exist, the negative refractive index can
be obtained. Detailed computational analysis is as follows.

\begin{figure}
\includegraphics[width=8cm]{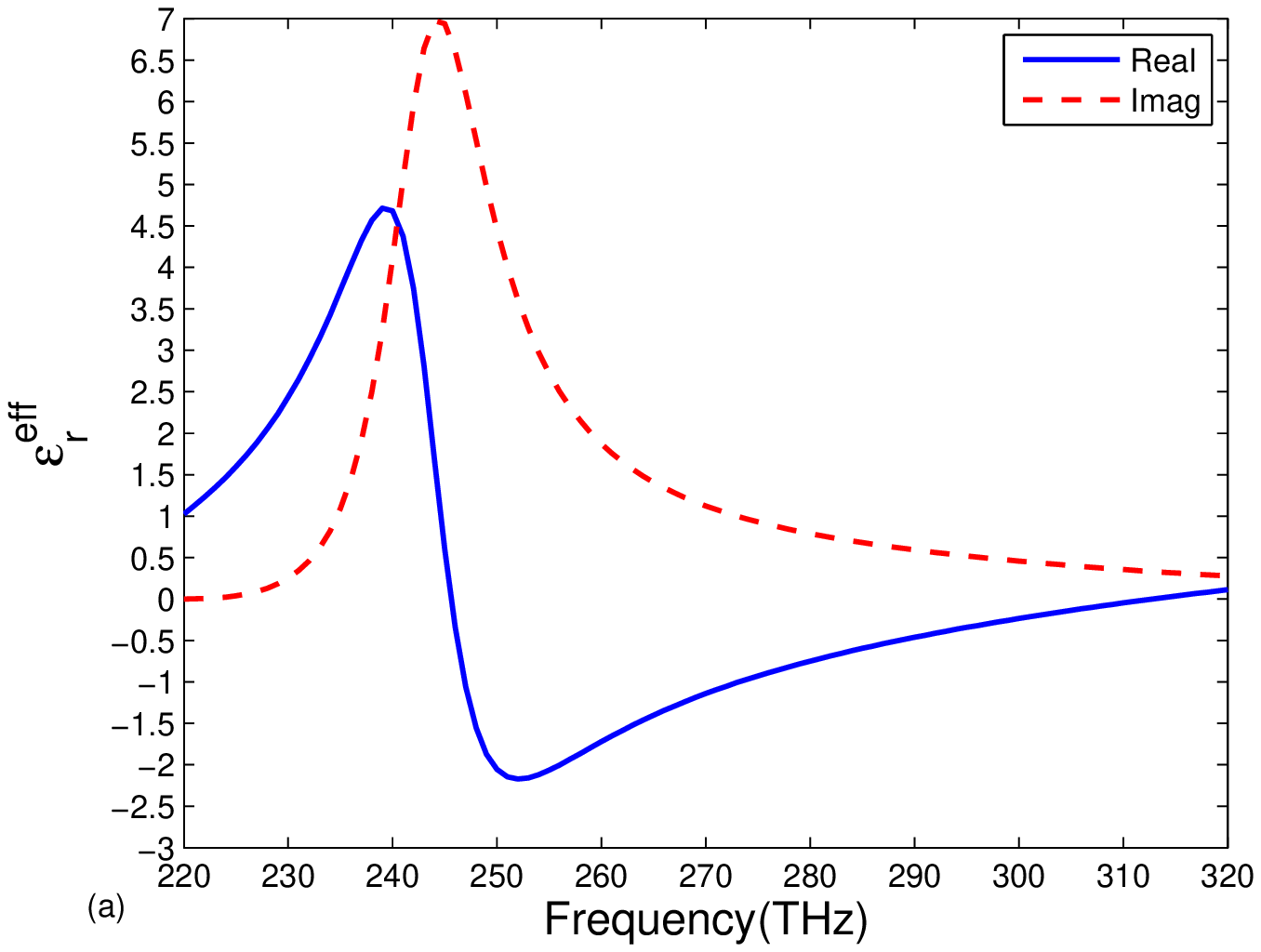}
\includegraphics[width=8cm]{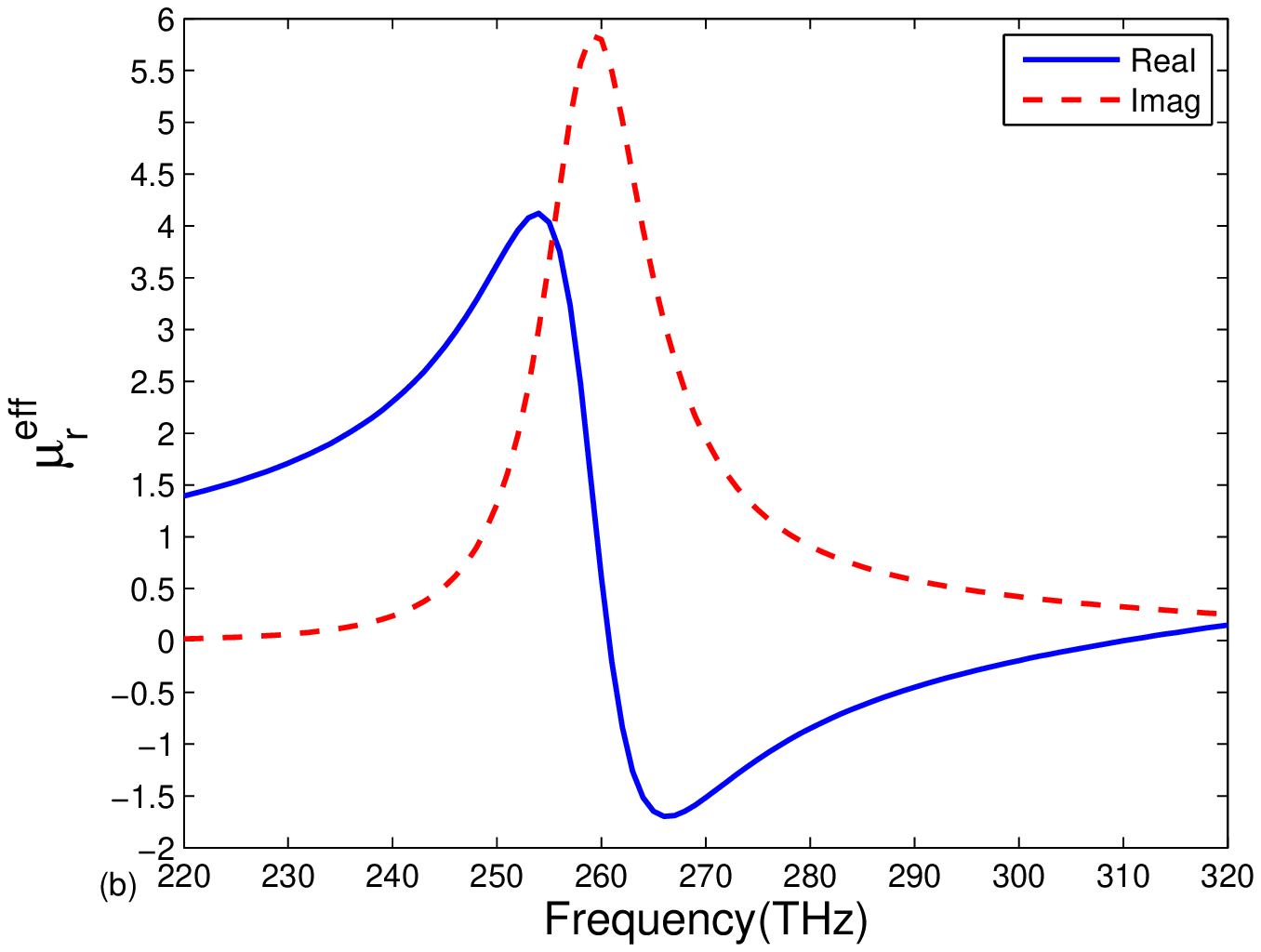}
\includegraphics[width=8cm]{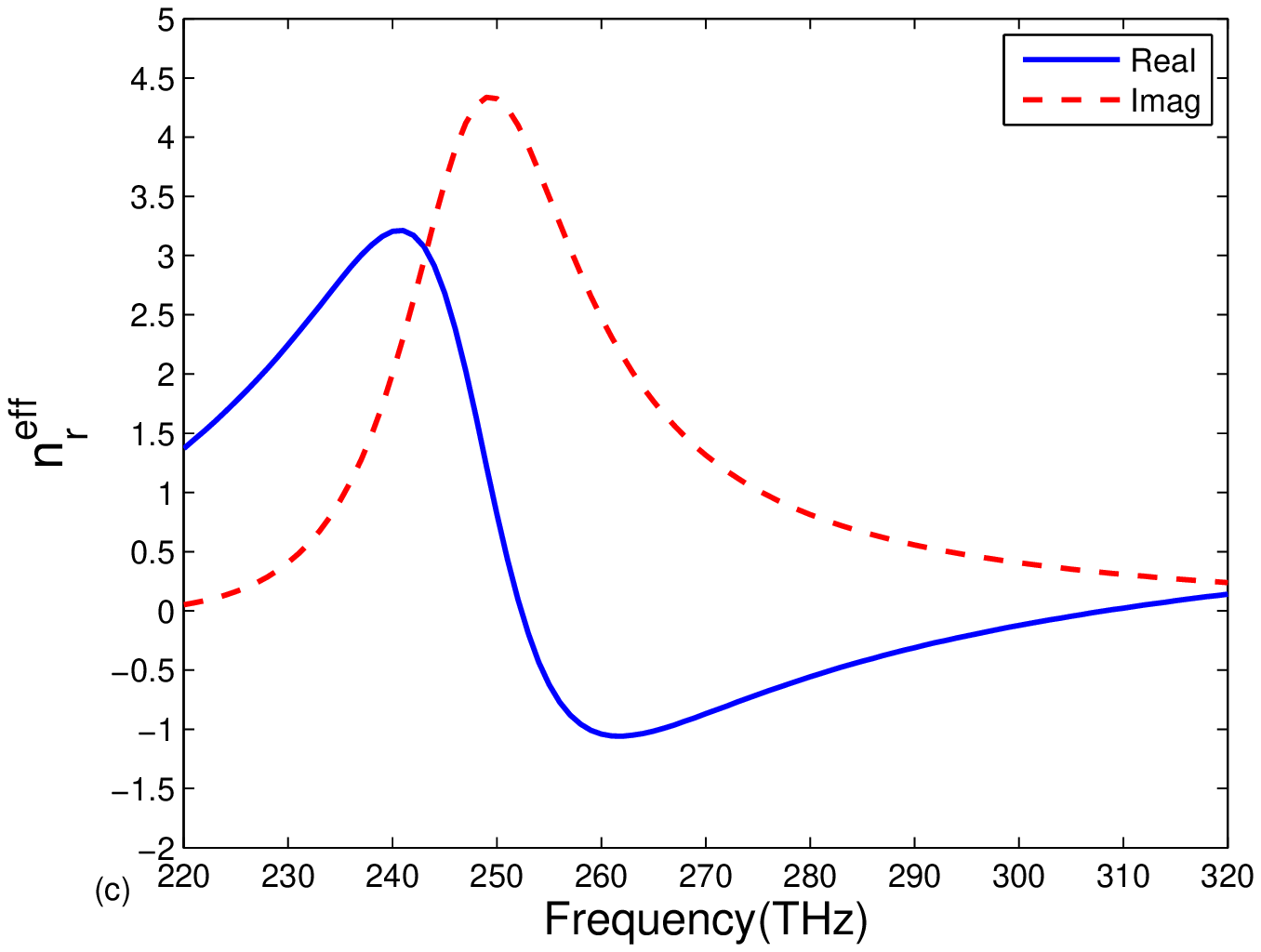}
\caption{\label{FuFig3}The effective permittivity, permeability and
refractive index of a collection of coated spheres. The dielectric
constant of the core is $\epsilon_1=2.2$ and the radius $
a_1=120nm$, and the coating is made of $\text{LiTO}_3$ with $
\epsilon_2=13.4$ and $a_2=130nm$. The dielectric constant of the
host medium is ${\bf \epsilon}_3=1.0 $ and the number density are
$N_\epsilon=N_\mu=(271.4nm)^{-3}$.}
\end{figure}

   According to the regulation of the Mie scattering coefficient, we
consider a collection of coated dielectric composite particles, with
the core (region 1) designed as Sec.\ref{II} A ($a_1=120nm,
\epsilon_1=2.2$) which provides $\mu_r^{eff}<0$, and with the shell
(region 2) designed as Sec.\ref{II} B ($a_2=130nm, \epsilon_2=13.4$)
which provides $\epsilon_r^{eff}<0$, embedded in a host medium
(region 3, $\epsilon_3=1.0$). The particle number density is fixed
to $N_\epsilon=N_\mu=(271.4nm)^{-3}$. This can be considered as a
periodic lattice of nanospheres with radius $a=130nm$ and
periodicity $R=480nm$. Scattering of Electromagnetic waves from
coated spheres has been worked out and $a_1$, $ b_1$ are written as
\cite{Bohren1983,Fung1994}
\begin{eqnarray}\label{h_+/e_+}
a_1=\frac{{\bf \psi}_1(x_2)-s_1{\bf \psi}'_1(x_2)} {{\bf
\xi}_(x_2)-s_1{\bf \xi}'_1(x_2)},
\end{eqnarray}
\begin{eqnarray}\label{h_+/e_+}
b_1=\frac{t_1{\bf \psi}_1(x_2)-{\bf \psi}'_1(x_2)} {t_1{\bf
\xi}_1(x_2)-{\bf \xi}'_1(x_2)},
\end{eqnarray}
respectively, where
\begin{eqnarray}\label{h_+/e_+}
s_1=n_2\frac{{\bf \psi}_1(n_2x_2)-p_1{\bf \chi}_1(n_2x_2)} {{\bf
\psi}'_1(n_2x_2)-p_1{\bf \chi}'_1(n_2x_2)},
\end{eqnarray}
\begin{eqnarray}\label{h_+/e_+}
t_1=n_2\frac{{\bf \psi}'_1(n_2x_2)-q_1{\bf \chi}'_1(n_2x_2)} {{\bf
\psi}_1(n_2x_2)-q_1{\bf \chi}_1(n_2x_2)},
\end{eqnarray}
\begin{eqnarray}\label{h_+/e_+}
p_1=\frac{n_2{\bf \psi}_1(n_2x_2){\bf \psi}'_1(n_1x_1)- n_1{\bf
\psi}'_1(n_2x_2){\bf \psi}_m(n_1x_1)} {n_2{\bf \psi}'_1(n_1x_1){\bf
\chi}_1(n_2x_2)-n_1{\bf \psi}_1(n_1x_1){\bf \chi}'_1(n_2x_2)},
\end{eqnarray}
\begin{eqnarray}\label{h_+/e_+}
q_1=\frac{ n_2{\bf \psi}'_1(n_2x_2){\bf \psi}_1(n_1x_1)- n_1{\bf
\psi}'_1(n_1x_1){\bf \psi}_1(n_2x_2)} {n_2{\bf \psi}_1(n_1x_1){\bf
\chi}'_1(n_2x_2)-n_1{\bf \psi}'_1(n_1x_1){\bf \chi}_1(n_2x_2)}.
\end{eqnarray}
Here $x_1=k_0r_1 $, $x_2=k_0r_2 $ and ${\bf \chi}_1(z)=-zy_1(z) $
where $y_1(z) $ is the spherical Bessel function of the second kind.
With the parameters stated above, magnetic resonant frequency is
predicted at 265Thz and dielectric resonant frequency is predicted
at 250THz by substituting $a_1$ and $b_1$ in Eq.~(\ref{g}) and
Eq.~(\ref{h}). To get the effective material values for coated
spheres, we simply substitute $a_1 $ and $b_1$ in Eqs.~(\ref{e}) and
(\ref{f}). The full calculations of the effective permeability and
permittivity are shown in Fig.~\ref{FuFig3}. These results coincide
with our prediction. The effective index is calculated with
$n_{eff}=\sqrt{\mu_r^{eff}\epsilon_r^{eff}}$ and ensuring
$n''_{eff}\geq0$. The maximum negative index is obtained with the
value about -1.2 at 260Thz where the magnetic and the dielectric
response simultaneously occur.

In the above, a periodic distribution of coated nanospheres is shown
to exhibit negative refractive index at optical frequencies. As for
random distribution of units, previous studies have shown that the
effective magnetic and electric resonances are different from
periodic structure. Specifically, Yannopapas has proven that
disordered distribution of elements can slightly affect values of
$\mu_r^{eff} $ and $\epsilon_r^{eff} $. However, the magnetic and
electric resonances behave qualitatively in the same way as in the
periodic case \cite{Yannopapas2006,Yannopapass2007}. It is because,
though the symmetry of the lattice can slightly influence the Mie
resonance, the determinant factors are the size, number density and
dielectric properties of the spheres (as shown in Eqs. (\ref{e}) and
(\ref{f})) \cite{Huang2004,Wheeler2006,Alu2005}. Therefore, even
nonperiodic distribution of coated spheres stated in this paper can
realize the NIMs at optical frequencies.

\section{Conclusion}
  In this paper we have theoretically shown a type of all dielectric
NIM at optical frequency by properly designing related parameters of
dielectric nanospheres. Near the frequencies of magnetic and
electric Mie resonances provided by the dielectric core and shell
respectively, both negative permeability and negative permittivity
are produced, and then left-handed metamaterials are obtained in the
optical domain. Note that $\text{LiTO}_3 $ coated nanosphere is just
taken as demonstration. Actually, due to the highly tunable
parameters of the all-dielectric concentric spheres, these proposed
structures enable a broad frequency range of negative refractive
index from deep infrared to visible domain. This is important if
such structures are to be used in practical NIM-based applications
at optical frequencies. Future work will be desirable to choose
appropriate dielectric materials for the coated spheres to fabricate
3D isotropic negative refractive index metamaterial in visible
frequencies.

\section*{Acknowledgements}
This work was supported in part by the National Natural Science
Foundation of China (No.~10847121,~10904036).




\end{document}